\documentclass{article}

\PassOptionsToPackage{super,comma,sort&compress}{natbib} 
\usepackage[preprint]{style}

\usepackage[utf8]{inputenc} 
\usepackage[T1]{fontenc}    
\usepackage{hyperref}       
\usepackage{url}            
\usepackage{booktabs}       
\usepackage{amsfonts}       
\usepackage{nicefrac}       
\usepackage{microtype}      
\usepackage{amsmath,amsthm,amsfonts,amssymb,amscd}
\usepackage{mathtools}
\usepackage{lastpage}
\usepackage{enumerate}
\usepackage{fancyhdr}
\usepackage{mathrsfs}
\usepackage[x11names]{xcolor}
\usepackage{graphicx}
\usepackage{listings}
\usepackage{subcaption}
\usepackage[english]{babel}
\usepackage{apxproof}
\usepackage{thmtools}
\usepackage{thm-restate}
\usepackage{enumitem}
\usepackage{float}
\usepackage{bm}
\usepackage{wrapfig}
\usepackage{algorithm}
\usepackage{algpseudocode}

\usepackage{sidecap}
\usepackage{siunitx}

\usepackage{outlines}

\theoremstyle{definition}

\definecolor{myGreen}{rgb}{0,0.5,0}

\makeatletter
\renewcommand{\normalsize}{%
  \@setfontsize\normalsize{13pt}{15.6pt}%
  \abovedisplayskip      9\p@ \@plus 2\p@ \@minus 5\p@
  \abovedisplayshortskip \z@ \@plus 3\p@
  \belowdisplayskip      \abovedisplayskip
  \belowdisplayshortskip 6\p@ \@plus 3\p@ \@minus 3\p@
}
\renewcommand{\section}{%
  \@startsection{section}{1}{\z@}%
                {-2.0ex \@plus -0.5ex \@minus -0.2ex}%
                { 1.5ex \@plus  0.3ex \@minus  0.2ex}%
                {\fontsize{17pt}{20.4pt}\selectfont\bfseries\raggedright}%
}
\renewcommand{\subsection}{%
  \@startsection{subsection}{2}{\z@}%
                {-1.8ex \@plus -0.5ex \@minus -0.2ex}%
                { 0.8ex \@plus  0.2ex}%
                {\fontsize{14.5pt}{17.4pt}\selectfont\bfseries\raggedright}%
}
\renewcommand{\subsubsection}{%
  \@startsection{subsubsection}{3}{\z@}%
                {-1.5ex \@plus -0.5ex \@minus -0.2ex}%
                { 0.5ex \@plus  0.2ex}%
                {\fontsize{13pt}{15.6pt}\selectfont\bfseries\raggedright}%
}
\makeatother
\normalsize

\title{More Electrodes, Faster Minds? Rethinking Bandwidth in Brain--Computer Interfaces}

\author{
  Boxuan Jiang \\
  School of Biomedical Engineering, Shanghai Jiao Tong University \\
  \texttt{data.j@sjtu.edu.cn}
}

\begin{document}

\maketitle

\section*{In Brief}
High-bandwidth BCIs can improve communication, control, sensory restoration,
and closed-loop therapy. This Perspective asks how these gains scale with
interface capacity. Orders-of-magnitude increases in meaningful human I/O face
constraints from the embodied organization of behavior, learning, and the
conditions under which neural states become subject-level expression.

\section{Abstract}
High-bandwidth brain--computer interfaces (BCIs) can bypass damaged pathways,
reduce motor costs, and improve communication and control. They also inspire
visions of accelerated thought output, mind reading, and instant skill
acquisition. This Perspective asks how gains in meaningful human I/O scale with
interface capacity. We distinguish bandwidth, decodable neural states,
neural states, and information a person can use, confirm, and express. Slowly
updated task states can unfold into complex behavior through the body, neural
control, sensory feedback, the environment, and shared context. Decodable neural
activity can support prediction and control;
subject-level communication depends on selection, confirmation,
and authorization. On the input side, stimulation may guide plasticity and
accelerate learning, while embodied skills arise through coordination of a
brain, body, and environment. The scaling relationship is likely nonlinear:
higher-capacity interfaces can yield real gains, while extreme increases in
meaningful human I/O encounter
constraints rooted in embodiment, learning, and subject expression.

\section{Introduction: How far can BCIs enhance human I/O?}

Public and industrial accounts of brain--computer interfaces (BCIs) often
follow a brain-centered input--output narrative. Natural channels such as the
hands and the vocal tract are cast as bodily intermediaries that constrain the
speed of human--machine communication, while a BCI promises more direct access
to the brain~\citep{wolpaw2002brain,serim2024revisiting,garbe2024presentation}.
This view is explicit in Neuralink's vision of increasing human--machine
bandwidth and achieving symbiosis with artificial intelligence, as well as in
Facebook's plan for high-speed typing ``directly from the brain''
~\citep{facebook2017braintyping,musk2019integrated}. Analyses of news and social
media further show that human enhancement occupies a prominent place in public
accounts of BCIs, with Neuralink and Elon Musk exerting a strong agenda-setting
effect~\citep{gilbert2019increasing,almanna2025public}.

These accounts share an intuitive model of a high-speed brain coupled to
low-bandwidth bodily peripherals. On this view, a person's internal life is far
richer than the content that can be conveyed through speech, typing, or other
overt behavior. An interface of greater bandwidth and lower latency can shorten
the path from an intention to an external action, especially when natural
channels are damaged, slow, or difficult to control. A stronger extrapolation
projects these gains across several orders of magnitude: as physical interface
bandwidth grows, the rates of meaningful human expression, reception, and
action should rise with it. We refer to this direct scaling assumption as the
\emph{BCI I/O myth}.

The same model supports aspirations ranging from direct thought reading and
high-speed language production to knowledge upload and skill implantation. On
the output side, the interface is expected to release thought from bodily
constraints; on the input side, sensations, knowledge, and skills are treated
as data that can be written into the nervous system. Both aspirations assume
that meaningful mental content already exists in a transmissible form.

The argument turns on what \emph{bandwidth} means. In BCI engineering, the term
broadly describes recording or stimulation capacity, including the number of
channels, electrode coverage, and spatial and temporal resolution. These
quantities characterize how many neural signals an interface can record or
perturb. Meaningful human I/O introduces two further levels: the discriminable
neural states that a system can extract or induce, and the rate at which a
person can use, confirm, express, or integrate those states. A recording system
may acquire an enormous volume of neural data while yielding little
task-relevant information. A stimulation system may perturb many neural states
at once, while stable perception, knowledge, and skill depend on how those
states enter an existing brain--body--environment system. Interface capacity,
neural information, and meaningful human I/O may therefore follow different
scaling curves. Early gains can come from bypassing damaged pathways, reducing
motor costs, improving signal estimation, and using feedback more effectively.
Larger gains increasingly depend on how quickly meaningful states are formed,
selected, confirmed, and integrated by the user.

\begin{figure}[t]
  \centering
  \includegraphics[width=\textwidth]{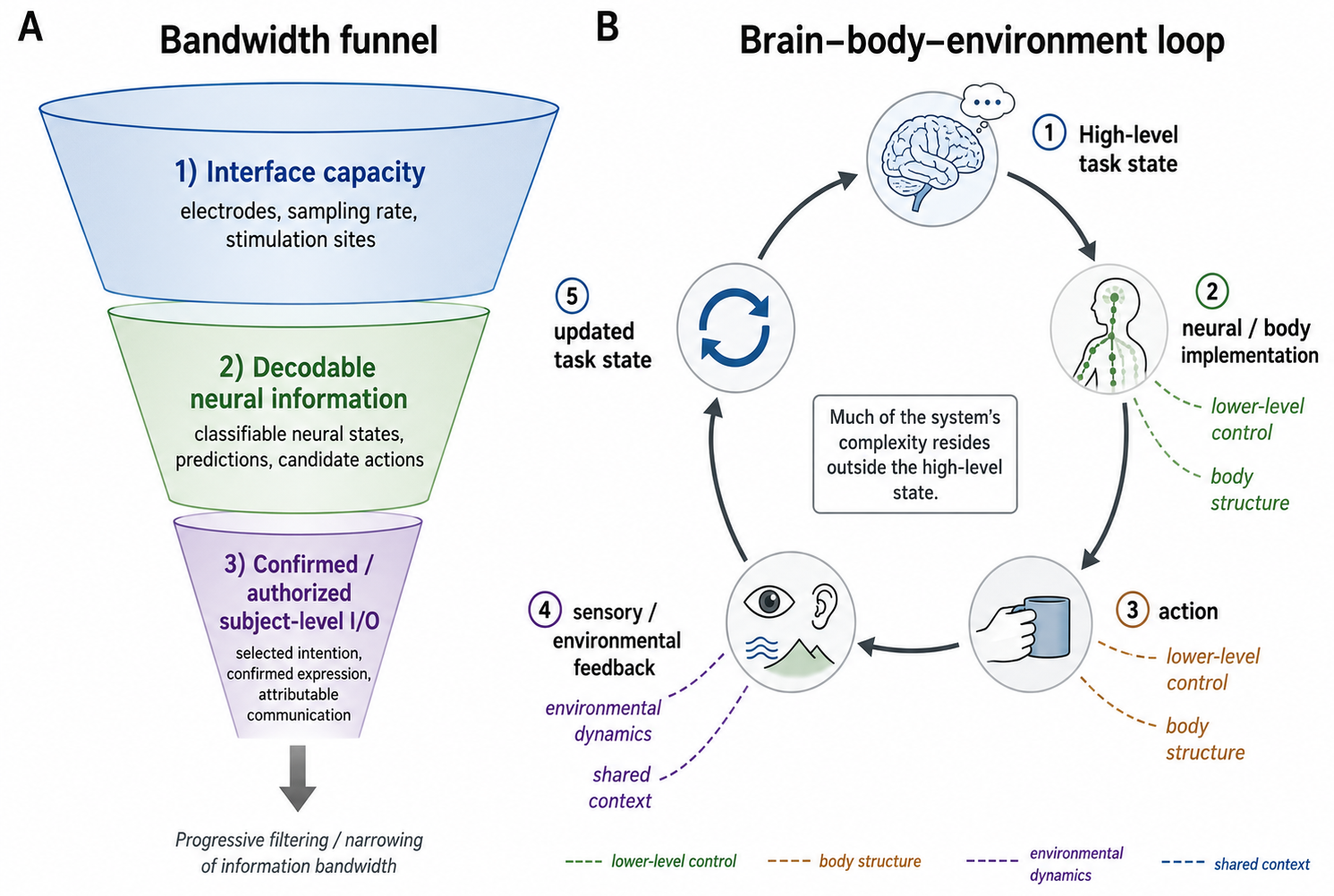}
  \caption{A framework for relating BCI bandwidth to meaningful human I/O.
  (A) The bandwidth funnel distinguishes physical interface capacity, decodable
  neural information, and confirmed or authorized subject-level I/O. Each level
  can support gains in the next, while the effective bandwidth narrows through
  estimation, selection, and confirmation. (B) A high-level task state is
  implemented through lower-level neural control, bodily structure, action,
  sensory and environmental feedback, and shared context. Much of behavioral
  complexity therefore resides in the recurrent brain--body--environment loop.}
  \label{fig:bandwidth-framework}
\end{figure}

Figure~\ref{fig:bandwidth-framework} summarizes the two linked parts of this
framework: information narrows as it moves from interface capacity toward
confirmed subject-level I/O, while complex behavior unfolds through a recurrent
brain--body--environment loop.

We therefore ask how far greater interface bandwidth can increase meaningful
human I/O, how those gains scale, and where strong diminishing returns may
emerge. Does the brain contain a stable, semantically organized stream of
high-level intentions whose rate far exceeds that of natural behavior? Can
embodied skills be acquired at radically higher speeds through sufficiently
precise stimulation?

We examine these questions through the organization of the full
brain--body--environment system. Complex behavior can be organized by slowly
updated high-level states and unfolded through bodily structure, lower-level
neural control, sensory feedback, environmental dynamics, and shared context.
This distribution of complexity offers an alternative account of how humans
produce rich behavior through a relatively slow stream of task-level choices.

We first consider the slow rate of task-relevant information flow in human
behavior and how low-dimensional states can organize high-dimensional action.
We then examine the two sides of the interface: when neural states read by a
high-bandwidth BCI can represent a user's expression, and whether neural
stimulation can write an embodied skill while bypassing training. Finally, we
consider the practical value of high-bandwidth BCIs for robust decoding,
sensory restoration, motor control, agency, and closed-loop neuromodulation.

\section{Where does behavioral complexity reside?}

\subsection{The slow rate of task-relevant human information flow}

Zheng and Meister recently surveyed information rates across a wide range of
human behaviors in their Perspective, \emph{The Unbearable Slowness of Being}
~\citep{zheng2025unbearable}. Their survey leads to a strikingly counterintuitive
observation. Typing, movement, perception, and memory rely on very different
sensory modalities, motor channels, and cognitive abilities. Yet the information
that actually changes the outcome of these tasks tends to flow at a similarly
slow rate, with a typical order of magnitude of about 10 bits/s.

What exactly does this rate measure? Given a task and a set of possible actions,
it asks how much uncertainty a person can eliminate per second about the
eventual behavioral outcome. Take typing. The rate can be estimated from the
number of characters produced per second and the entropy of the character
sequence. Pressing a different key changes the outcome and counts as
information. Pressing the same key slightly harder, blinking between two
keystrokes, or producing an unrelated muscle twitch does not change the text.
None of these variations increases the task-relevant information rate.

In this article, \emph{task-relevant information flow} refers to the rate at
which a person can form, select, and turn such distinctions into behavioral
outcomes. It captures the update speed of high-level behavioral states: how many
possible actions a person can choose among at a given moment, and how quickly
those choices become visible in language, tool use, or other actions.

These observations constrain how gains from higher interface bandwidth should
be interpreted. More electrodes can improve signal estimation, reduce errors,
and shorten the path from an intention to an external action. The formation of
the next task-relevant state may scale more slowly. Zheng and Meister capture
this distinction half in jest: for language communication, typing through an
ultra-high-channel-count Neuralink may eventually approach the same task-level
rate as making a phone call, even as the interface becomes faster and more
reliable.

Existing estimates come from behavior produced through a natural body, a
natural output channel, and a fixed task interface. These measurements leave
room for gains from bypassing peripheral bottlenecks and for the possibility of
a faster internal stream of plans or semantic content. The magnitude of that
latent stream remains an empirical question. The following sections examine how
the brain--body--environment system may shape its scaling. Much of behavioral
complexity can reside in bodily structure, lower-level neural control, sensory
feedback, environmental constraints, and shared context.

\subsection{The brain never evolved without a body}

Evolution never had the option of designing a brain first and attaching a body
later. The nervous system has always belonged to an organism that must survive
in a physical environment. Its basic problem is therefore a closed-loop one:
to perceive, plan, and act well enough to keep that organism alive. Adaptive
behavior emerges from the interaction of the nervous system with a particular
body in a particular world~\citep{chiel1997brain}.

Perception, planning, and action are tightly coupled within this loop. During
an eye movement, for example, a corollary discharge informs sensory circuits
about the movement being generated, allowing perception to account for the
consequences of the animal's own action~\citep{crapse2008corollary}. Buzs{\'a}ki
pushes this relation further: an organism first had to become an actor before
it could become a sophisticated perceiver, and the original value of perception
was to guide action~\citep{buzsaki2019brain}. An organism capable of action with
little perception may still stumble into survival. An organism that perceives
beautifully and cannot act has a much shorter career.

The evolution of perception and planning therefore took place under persistent
constraints from the action system. A nervous system must learn to control the
body it actually has: a body with particular conduction delays, muscles,
joints, sensory organs, and dynamics. These properties shape which control
signals are useful, how quickly feedback arrives, and which corrections remain
possible. In this sense, the body participates in the design problem faced by
the brain~\citep{chiel1997brain}.

A striking example suggests that these bodily constraints reach into the
temporal organization of the brain itself. Mammalian brain volume increased by
several thousandfold over evolution, yet the hierarchy of brain rhythms has
remained remarkably conserved across species~\citep{buzsaki2013scaling}.
Neocortical alpha rhythms, sleep spindles, and hippocampal ripples show similar
frequencies, temporal profiles, and waveforms in animals with vastly different
brain sizes. Brain size scaled. Timing largely did not.

This conservation creates an engineering problem. A larger brain contains
longer communication paths, which should increase delays between distant
regions. Buzs{\'a}ki, Logothetis, and Singer argue that a small population of
large-diameter, fast-conducting long-range axons may compensate for the added
distance~\citep{buzsaki2013scaling}. Buzs{\'a}ki later offers a more provocative
possibility: the body's effectors may help set the clock that the brain must
keep~\citep{buzsaki2019brain}. The basic properties of myosin and actin are
broadly conserved across mammals. Motor commands generated by the motor cortex,
cerebellum, and basal ganglia must therefore arrive within comparable temporal
windows. The physical timescale of the action system remains comparatively
stable as brain size increases, and the neural systems controlling it preserve
their coupling to that timescale.

This observation provides an evolutionary intuition for the scaling of BCI
gains. The temporal organization of the brain is deeply embedded in the physical
constraints of the body. Human perception, planning, and action developed
around a body operating at its present timescale. Modest improvements can arise
from more efficient access to this system. An intention stream operating orders
of magnitude faster than embodied behavior would require temporal organization
beyond the regime for which the system evolved. The mouth and hands may look
like slow peripherals from the viewpoint of a computer interface, yet the brain
that controls them has never lived on the other side of that separation.

\subsection{Low-dimensional states and high-dimensional behavior}

Zheng and Meister's analysis leaves a further question open: why do typing,
movement, memory, and other complex tasks all yield such low information rates
~\citep{zheng2025unbearable}? The measure assigns additional information to
effective updates of a high-level task state. Continuous changes during
execution contribute additional bits when they alter the space of behavioral
choices.

Consider turning a Rubik's cube. At a given moment, a skilled solver may face
only a few high-level decisions: which face to turn next, which learned sequence
to use, or whether to change the current strategy. These choices may be updated
only a few times per second. Yet a seemingly simple turn involves finger
positioning, grip stabilization, muscle coordination, joint movement, tactile
correction, visual confirmation, and use of the cube's mechanical structure.
These processes have high dimensionality and fine temporal precision, but they
all serve the same high-level goal and are not counted separately in the
task-relevant information rate.

The high-level state therefore needs to maintain only a relatively
low-dimensional control variable, such as ``make this turn'' or ``continue the
current solution.'' Once formed, this goal can unfold into a sequence of
continuous and precise movements through trained motor programs, spinal and
cerebellar control, sensory feedback, and the physical constraints of the cube.
The behavior is complex, while the content that must be repeatedly selected at
the high level remains small.

Typing and language have a similar structure. Deciding what to express next
provides much of the task-relevant information. The movement of the fingers
toward the keys, visual correction of their position, and execution of practiced
movement sequences are handled by an existing brain--body--environment control
system~\citep{chiel1997brain}. Under Zheng and Meister's measure, these execution
processes contribute through the high-level choice they realize. Language also
relies on a much larger external background. A short sentence can convey rich content
because both speakers already share a language, a conversational history, an
object of attention, a social relationship, and a surrounding environment. The
speaker only needs to provide a small amount of information that indexes and
updates this shared context. A few words can call upon a large structure of
knowledge; the transmitted increment can be small even when the resulting
understanding is rich.

From this perspective, approximately 10 bits/s appears less unusual. The measure
captures the slowest and most compressed part of the full system: high-level
goals, behavioral choices, and contextual updates. A low-dimensional goal can
unfold through the body, spinal cord, senses, and environment into behavior that
is high-dimensional, continuous, and fast. Bodily structure, lower-level neural
control, practiced actions, sensory feedback, environmental dynamics, and
shared background carry much of this complexity without reappearing in the
task-relevant information rate.

A BCI can make the execution and readout of these goals faster and more
reliable. Further scaling increasingly depends on the rate at which the
high-level goals themselves are formed and updated.

\section{The output side: what does a high-bandwidth BCI read?}

The previous section proposed that slowly changing high-level states can
organize complex behavior by recruiting the body, lower-level neural control,
environmental feedback, and shared context. Greater recording bandwidth gives
a BCI a wider and sharper view of the neural activity involved in this process.
The central question now shifts. As the interface gains access to more neural
states, which of those states can speak for the user?

\subsection{Neural decodability and subject expression}

The image of a high-speed brain constrained by low-bandwidth bodily peripherals
continues a distinctly dualist model of mind. Genuine thoughts, intentions, and
meanings are assumed to form inside the brain, with the body serving as their
outlet to the external world.

A full treatment of mind--body dualism lies beyond the present discussion. We
still need to understand why this intuition has such a persistent hold on how
people imagine BCIs.

In everyday experience, it is easy to feel that we possess a subject located
somewhere ``inside'': a subject who thinks, chooses, and expresses through the
body. Social institutions rely on a similar model. Law, ethics, and ordinary
interaction rarely require a complete ontology of neural activity; they rely on
intention, consent, expression, attribution, and responsibility. Human society
therefore needs an \emph{operational subject interface}: content must be
recognizable to the subject, confirmed as its expression, and attributable to
it before entering ordinary communication and systems of responsibility
~\citep{haggard2017agency}.

This argument requires no central interpreter or Cartesian homunculus assigning
meaning inside the brain. The self or conscious will may be a virtual machine,
an illusion, or any of the other metaphors favored by consciousness scientists
and philosophers. The operational requirement remains. Moreover, the subject
state available for explicit expression may update at a limited rate because
active choices, inner speech, and conscious reports are constrained by
attention, working memory, and serial processing. The brain contains enormous
simultaneous activity, while relatively little of it enters the subject
interface as explicit expression.

Rich neural activity offers many signals for prediction and control
~\citep{haynes2006decoding}. Its conversion into subject-level communication
introduces an additional operation: the user must select, recognize, and
authorize a decoded state as an expression. This operation also gives the state
a place within ordinary social and ethical structures
~\citep{yuste2017ethical}.

\subsection{High-bandwidth BCIs as side-channel probes}

As recording bandwidth increases, a BCI can read sensory processing, motor
preparation, action candidates, error signals, attentional biases, emotion and
arousal, fatigue, sensory predictions, and the unresolved competition between
different cognitive and action states with increasing precision. These signals
can provide important information for behavioral decoding. A system may predict
that a user is about to reach, speak, or select a target before the action
occurs. It may also identify a word being prepared, a stimulus receiving
attention, or an emotional state that has not entered explicit report
~\citep{haynes2006decoding,cisek2010action}.

In this sense, a high-bandwidth BCI resembles a side-channel probe. By observing
states inside the brain, it obtains information related to the user's cognition
and behavior. Neural decodability, behavioral prediction, communicative
intention, and subject confirmation nevertheless occupy different levels. A
word may be activated in the language system while remaining only one of
several candidate expressions. An action may enter preparation and later be
inhibited or cancelled. An emotion or preference may be identified by a
classifier before it becomes a position that the user is willing to express
publicly.

Speech BCIs make this distinction especially clear. A system must distinguish
language that the user is preparing to say from language that is being read
silently, imagined, actively suppressed, or incidentally activated. Recent
systems can decode attempted speech at high rates and reconstruct semantic
content from both perceived and imagined language
~\citep{willett2023speech,tang2023semantic}. Predictive decoding can reduce the
number of explicit selections, anticipate intended words, and correct errors
before overt output. These mechanisms can produce real gains in effective
communication rate. As decoding reaches further into imagined, suppressed, and
incidentally activated language, user selection and confirmation determine
which states enter the external channel.

Increasing recording bandwidth can enlarge both the information available for
prediction and the set of states awaiting subject confirmation. Early gains
come from better estimation and fewer explicit selections. At larger scales,
the rate of authorized expression becomes an increasingly important part of the
system's effective bandwidth ~\citep{yuste2017ethical}.

\section{The input side: How far can BCIs accelerate embodied skill acquisition?}

High-bandwidth stimulation may provide richer feedback, guide plasticity, and
accelerate learning. A stronger input-side vision treats sensations, knowledge,
skills, and memories as files that can be written into the nervous system,
allowing a complete ability to appear with little training or calibration. We
refer to this direct transition from stimulation bandwidth to instant ability
as the \emph{input myth}.

The possibility of writing purely semantic knowledge through a neural interface
leaves room for a broader discussion. Here we set that question aside and focus
on embodied skills, for which the difficulty of direct upload is already clear.
Such skills occupy a substantial part of human ability. Riding a bicycle,
swimming, playing an instrument, writing, driving, playing sports, and using
tools all depend on sensorimotor coordination acquired over time. These forms of
``skill knowledge'' are realized through continued coupling among the brain,
spinal cord, muscles, joints, sensory feedback, and environmental constraints
~\citep{chiel1997brain}.

Consider piano playing. Skilled performance involves finger independence,
muscle coordination, keystroke force, rhythmic control, auditory prediction,
visual positioning, error correction, and motor programs built through long
practice. Immediate transfer of a cortical activity pattern from one pianist to
another would require that pattern to match the recipient's hand structure,
muscle control, sensory calibration, prior experience, and error-correction
dynamics at the same time.

The same point applies to riding a bicycle. Balance is a real-time coordination
problem spanning vision, vestibular sensation, proprioception, body posture,
bicycle dynamics, and environmental feedback. During learning, a person
repeatedly produces an error, senses the error, and corrects it. Through this
process, the full system gradually enters a more stable state of coordination
~\citep{wolpert1995internal,krakauer2019motor}.

An embodied skill is therefore closer to a stable dynamical organization formed
through training of the brain--body--environment system. Learning changes neural
representations together with sensory predictions, motor control, bodily
calibration, error feedback, and adaptation to the environment. BCIs may
accelerate this process through richer feedback, guided plasticity, and faster
error correction ~\citep{sadtler2014constraints}. Complete skill acquisition
additionally requires calibration across muscle control, proprioception, body
dynamics, and the environment.

Skill acquisition thus involves two levels: the neural state observed after
training and the learning process that brings the full system into stable
coordination. The first may be recorded, stimulated, or partly reconstructed.
The second reorganizes a particular brain, body, and environment over time. The
distance between accelerated learning and instant upload is the distance between
modifying a neural state and establishing this coordinated dynamical
organization. The direct writing of semantic knowledge, factual memory, and
abstract concepts remains a separate question.

\section{Conclusions and outlook: What can high-bandwidth BCIs actually provide?}

Increasing recording bandwidth, stimulation precision, and closed-loop speed
remains one of the most valuable directions for BCI research and deserves
sustained effort ~\citep{lebedev2017brainmachine}.

Current systems retain major limitations in large-scale recording stability,
precision, and simultaneous stimulation. Practical bottlenecks arise from
limited recording coverage, low signal-to-noise ratios, poor long-term
stability, restricted stimulation degrees of freedom, and excessive closed-loop
latency. Increasing electrode
counts, expanding recording montages, improving spatial and temporal
resolution, maintaining stable recordings over time, and developing finer
stimulation and feedback systems therefore have clear technical and clinical
motivation.

Reliable recovery of a low-rate task variable may require neural recordings
with vastly greater raw bandwidth. Task-relevant states are embedded in noisy,
dynamic, and widely distributed activity. Richer recordings can therefore
improve accuracy, robustness, and generalization even when the final language
or control stream carries only tens of bits per second. Recent
high-performance handwriting and speech neuroprostheses illustrate what richer
recordings and decoding can yield at the behavioral interface
~\citep{willett2021handwriting,willett2023speech}.

For people with paralysis, limb loss, or impaired movement, high-bandwidth BCIs
can provide a more continuous, natural, and controllable interface for action.
A low-bandwidth system may distinguish only a few discrete commands---left,
right, click, or stop. A higher-bandwidth system can estimate trajectories,
speed, force, grasp configuration, movement preparation, and error signals,
while real-time feedback allows the user to correct the control continuously.
Neural control of reach and grasp already shows the clinical importance of
moving beyond a small command vocabulary toward continuous action
~\citep{hochberg2012reach}.

In neural prostheses, exoskeletons, and functional electrical stimulation, an
equally important goal is to restore agency: the experience that ``I am
controlling this,'' that ``this movement belongs to me,'' and that ``I can
correct it through feedback.'' Stable closed-loop control, natural sensory
feedback, and a recovered sense of bodily ownership have enormous clinical
value in their own right ~\citep{haggard2017agency}. For sensory restoration,
greater bandwidth enables finer stimulation and more precise shaping of
perception. More channels, richer stimulation parameters, lower latency, and
more stable feedback can establish a more accurate map within the user's
subjective perceptual space. In tactile and proprioceptive restoration, such
stimulation may turn feedback from a crude cue into a bodily sensation that
genuinely participates in action control; intracortical tactile feedback has
already been shown to improve robotic-arm control ~\citep{flesher2021tactile}.

For neurological disorders, higher-bandwidth recording and stimulation can
support finer recognition and regulation of brain states. Many disorders
involve abnormal rhythms, pathological synchrony, failed network-state
transitions, or disrupted dynamic regulation. Higher-resolution closed-loop
systems may detect pre-seizure states, Parkinsonian oscillations, and other
patterns requiring state-dependent intervention, then deliver stimulation at a
more appropriate time and with more appropriate parameters. Responsive
cortical stimulation for epilepsy and adaptive deep-brain stimulation for
Parkinson's disease provide concrete examples of this logic
~\citep{morrell2011responsive,little2013adaptive,gilron2021streaming}.

The relationship between interface bandwidth and meaningful human I/O is likely
nonlinear. Early gains can be substantial because BCIs can bypass damaged
pathways, reduce motor costs, improve prediction, and stabilize closed-loop
control. As the desired gain grows, the limiting factor increasingly shifts
toward the formation, confirmation, and embodied integration of meaningful
states. Subject expression requires content that a person wishes to express,
can confirm, and is willing to take responsibility for
~\citep{yuste2017ethical}. Orders-of-magnitude thought output, complete mind
reading, and instant skill implantation lie in this latter regime. The direct
writing of semantic knowledge, factual memory, and abstract concepts remains
beyond the present analysis.

BCIs may allow people to move again, feel again, and regain control over their
bodies and environments. They may also let us enter the
brain--body loop with unprecedented precision. At the present stage of BCI
development, we are closer to swimmers in open water than travelers approaching
land. Greater bandwidth is a flotation device: it helps us rise, breathe,
regain control, and keep moving. Its value is immediate and may be lifesaving.
The shore---frictionless thought transfer, complete mind reading, and instant
skill upload---remains a different destination. A flotation device can save a
life. Reaching the shore is the next and much harder problem.

\section*{Declaration of Interests}
The authors declare no competing interests.

\bibliographystyle{vancouver}
\bibliography{draft/ref}

\end{document}